\documentstyle[12pt,epsfig]{article}

\catcode`@=11
\newcount\@tempcntc
\def\@citex[#1]#2{\if@filesw\immediate\write\@auxout{\string\citation{#2}}\fi
  \@tempcnta\z@\@tempcntb\m@ne\def\@citea{}\@cite{\@for\@citeb:=#2\do
    {\@ifundefined
       {b@\@citeb}{\@citeo\@tempcntb\m@ne\@citea\def\@citea{,}{\bf
?}\@warning
       {Citation `\@citeb' on page \thepage \space undefined}}%
    {\setbox\z@\hbox{\global\@tempcntc0\csname b@\@citeb\endcsname\relax}%
     \ifnum\@tempcntc=\z@ \@citeo\@tempcntb\m@ne
       \@citea\def\@citea{,}\hbox{\csname b@\@citeb\endcsname}%
     \else
      \advance\@tempcntb\@ne
      \ifnum\@tempcntb=\@tempcntc
      \else\advance\@tempcntb\m@ne\@citeo
      \@tempcnta\@tempcntc\@tempcntb\@tempcntc\fi\fi}}\@citeo}{#1}}
\def\@citeo{\ifnum\@tempcnta>\@tempcntb\else\@citea\def\@citea{,}%
  \ifnum\@tempcnta=\@tempcntb\the\@tempcnta\else
   {\advance\@tempcnta\@ne\ifnum\@tempcnta=\@tempcntb \else
\def\@citea{--}\fi
    \advance\@tempcnta\m@ne\the\@tempcnta\@citea\the\@tempcntb}\fi\fi}
\catcode`@=12

\begin{document}
\title{\vskip-3cm{\baselineskip14pt
\centerline{\normalsize DESY 00-092\hfill ISSN 0418-9833}
\centerline{\normalsize hep-ph/0009111\hfill}
\centerline{\normalsize August 2000\hfill}}
\vskip1.5cm
Next-to-leading order Calculation \\ of a Fragmentation Function
\\ in a Light-Cone Gauge }
\author{{\sc Jungil Lee}\footnote{E-mail: jungil@mail.desy.de}\\
{\normalsize II. Institut f\"ur Theoretische Physik, Universit\"at
Hamburg,}\\
{\normalsize Luruper Chaussee 149, 22761 Hamburg, Germany}}

\date{}

\maketitle

\thispagestyle{empty}

\begin{abstract}
Next-to-leading order corrections to fragmentation functions in a 
light-cone gauge are discussed.
This gauge simplifies the calculation by eliminating many Feynman diagrams
at the expense of introducing spurious poles in loop integrals.
As an application, the short-distance coefficients for the color-octet 
$^3S_1$ term in the fragmentation function for a gluon to split into 
polarized heavy quarkonium states are re-calculated to order $\alpha_s^2$.
We show that the ill-defined spurious poles cancel and the appropriate 
prescriptions for the remaining spurious poles can be determined by 
calculating a subset of the diagrams in the Feynman gauge.
Our answer agrees with the recent calculation
of Braaten and Lee in the Feynman gauge, but disagrees with another previous
calculation.

\medskip

\end{abstract}

\newpage


\section{Introduction}
Factorization theorems for inclusive single-hadron production
\cite{C-S} guarantee that the dominant mechanism for a hadron production
with high $p_T$ is {\it fragmentation} \cite{FM}, 
the production of a parton which
subsequently decays into the hadron and  other partons.
Many of the theoretical uncertainties disappear in the high $p_T$ region
and  fragmentation is, therefore, a nice probe of  
the hadron production mechanism.
This process is described by a fragmentation function $D(z,\mu)$, where
$z$ is  the longitudinal momentum fraction of the hadron and $\mu$ is
a factorization scale.

 The earliest calculations of fragmentation functions exploited the fact 
that the fragmentation function is independent of the production processes 
of the decaying parton.  For example, the fragmentation functions 
for heavy quark\cite{M-N} and for  quarkonia 
\cite{B-Y:S1,B-C-Y:cS,B-C-Y:Bc} were deduced by comparing the production
cross sections for the hadron with the form predicted by the factorization
theorems for inclusive single-hadron production.
A field-theoretical definition of the fragmentation function
can be expressed as matrix elements of bilocal operators in a light-cone gauge
\cite{C-F-P} or, more generally, as matrix elements of nonlocal
gauge-invariant operators \cite{C-S}. 
The Collins-Soper definition of the fragmentation function was first used 
by Ma to calculate a fragmentation function for heavy quarkonium 
in leading order \cite{Ma-1} and in next-to-leading order \cite{Ma-2}. 
The definition is particularly convenient for carrying out calculations 
beyond leading order in $\alpha_s$, and it allows for the calculation 
in the Feynman gauge.  By using the Feynman gauge, one can avoid the problem 
caused by the ambiguity of the spurious pole of the gluon propagator 
in the light-cone gauge. On the other hand, one 
must calculate a number of diagrams which do not appear in the 
light-cone gauge. In higher-order corrections to a fragmentation function,
the simplicity in the light-cone gauge is remarkable, provided the spurious
poles are handled correctly.  The spurious pole problem does not appear
in  tree-level real-parton corrections.

However, when we calculate a virtual correction to an amplitude in the 
light-cone gauge, we should keep in mind the possible problem caused 
by the spurious pole.
A naive way to evaluate such light-cone dependent integrals is known as
the Cauchy principal value (PV) prescription.
If we use the PV prescription, there exist the ill-defined spurious 
pole in some loop integral. Without introducing {\it ad hoc} assumptions,
these integrals are not calculable using dimensional regularization. 
An elegant method has been presented by
Mandelstam and Leibbrandt independently\cite{ML-M,ML-L}.\footnote{
Comprehensive reviews can be found in Refs. \cite{REVIEW,BOOK1,BOOK2}.}
The ML prescription made it possible to transform such an integral 
into a well-defined one. 
The derivation of the Altarelli-Parisi evolution of parton densities
\cite{AP} is one of the best examples of the use of the light-cone gauge.
All previous calculations beyond-leading order had to employ some 
prescription for the spurious pole.
The PV prescription is used in Refs. \cite{C-F-P,C-F-P-2,AP-2-PV}.
The ML prescription is  employed in the leading order \cite{AP-1} and
the next-to-leading order \cite{AP-2-1,AP-2-2}.

A safe way to fix the prescription for the spurious pole is to find the gauge 
transformation relation from another gauge, such as the Feynman gauge, 
where all the poles are well defined.  
Fortunately, at least for the case of the 
fragmentation function calculation, one may make use of the gauge-invariant
property of the fragmentation function of Collins and Soper.
One notices that the spurious pole in the light-cone gauge 
is transformed into the propagator of the eikonal operator in the Feynman gauge 
which makes the nonlocal operators of the decaying partons, quark or gluon,  
gauge-invariant.  Therefore, it is not necessary to introduce 
a prescription for the spurious pole structure. 
Instead, by matching the result from the light-cone gauge, where the spurious 
pole has an ambiguity, with the well-defined Feynman gauge 
result, one can determine the structure by virtue of gauge invariance.

In this work, we develop a way to calculate next-to-leading 
order corrections to a fragmentation function in the light-cone gauge.
Instead of employing any known prescription for the spurious pole, 
we determine the sign of $i\epsilon$ in the spurious pole by using only 
gauge invariance. It is fixed by comparing the light-cone gauge result 
with that from another gauge where there is no such ambiguity.
This method, which has not been used before, is derived straight-forwardly 
from the gauge independent definition of the fragmentation function given 
by Collins and Soper \cite{C-S}.
By choosing the gauge-fixing vector $n$ with vanishing transverse component
with respect to the momentum of the produced hadron,
we observe that the ill-defined spurious poles in one-loop integrals disappear,
at least in our example. The reason is that the transverse momenta of the final 
states are integrated out. 
As an application, we re-derive the short-distance coefficients for 
the color-octet $^3S_1$ term in the fragmentation function for a gluon 
to split into polarized heavy quarkonium states to order $\alpha_s^2$.
There are two previous calculations of this function which disagree
with each other \cite{Ma-2,B-L}.
We use both the Feynman gauge and the light-cone gauge.
We fix the sign of $i\epsilon$ in the spurious pole by matching the two
calculations. Our results from the two gauges agree with each other 
before the evaluation of the loop integrals.  We remove ultraviolet divergences 
using the $\overline{\rm MS}$ renormalization procedure.
Our result agrees with the recent calculation of Braaten and Lee \cite{B-L}.

This paper is organized as follows.
In section \ref{sec:LC}, we give a short description of 
the spurious pole problem in the light-cone gauge, evaluate the
one-loop renormalization constants in the light-cone gauge and compare
our results with those based on the PV and the  ML prescriptions.
In section \ref{sec:CS}, we describe the method based on the Collins-Soper
definition for the calculation of a fragmentation function
in the light-cone gauge.
As an application of our method, we give in section \ref{sec:CAL}
the result for the color-octet $^3S_1$ term in the fragmentation function 
for a gluon to split into heavy quarkonia.
A discussion is presented in section \ref{sec:DIS}.
\section{One-loop correction in the light-cone gauge}
\label{sec:LC}
The light-cone gauge is a physical gauge where the gluon field $A^\mu$ has
vanishing light-cone projection
\begin{eqnarray}
A\cdot n=0,
\end{eqnarray}
where $n$ is an arbitrary light-like vector ($n^2=0$)
appearing in the gauge-fixing term in the QCD Lagrangian.
In the light-cone gauge, 
in which the fragmentation function was originally defined \cite{C-F-P},
the eikonal line as well as the ghost decouples from the gluon, since the
coupling, proportional to $n^\mu$, is orthogonal to the gluon 
propagator. One draw-back of the light-cone gauge in higher-order calculations
is the existence of the spurious pole $1/k\cdot n$ in the gluon propagator
\begin{eqnarray}
\frac{i}{k^2+i\epsilon}\left[
-g^{\mu\nu}+\frac{k^\mu n^\nu+n^\mu k^\nu}{k\cdot n}
\right],\; \epsilon>0,\; \mu,\nu=0,1,2,3,
\label{Gpro:LC}
\end{eqnarray}
where $k$ is the momentum of the gluon.
In order to evaluate a loop integral depending on the spurious pole,
one need to introduce some assumptions. 
Various prescriptions for the spurious pole have been proposed.
As a naive way, the PV prescription assumes
\begin{eqnarray}
\frac{1}{k\cdot n}\equiv \frac{1}{2}\left[\frac{1}{k\cdot n+i\epsilon}
                     +\frac{1}{k\cdot n-i\epsilon}\right].
\label{PV}
\end{eqnarray}
An elegant method is known as the ML prescription \cite{ML-M,ML-L}.
In this prescription, the spurious pole is re-expressed as
\begin{eqnarray}
\frac{1}{k\cdot n}\equiv
\lim_{\epsilon\to 0^+}
\frac{k\cdot \bar{n}}{k\cdot n \;k\cdot \bar{n}+i\epsilon},
\label{ML}
\end{eqnarray}
where $\bar{n}$ is a conjugate light-like vector
satisfying $\bar{n}^2=0$. Its spatial components are opposite in sign
to those of $n$.  The merit of the ML prescription is that it allows us 
for a proper Wick rotation to evaluate the integral in Euclidean space.
In the ML prescription, the reduction of tensor integrals into scalar integrals,
and the evaluation of scalar integrals, can be very involved due to the 
introduction of one more light-like vector $\bar{n}$.
The reason is that the light-cone vector has a non-vanishing
transverse component compared to that of the propagator momentum.

Let us classify the integrals involving spurious poles into two classes
in view of the naive PV prescription which does not include any other
assumption but (\ref{PV}).
First, there is an integral with the {\it ambiguous} spurious pole.
This integral can be regularized using dimensional regularization 
imposing the PV prescription  (\ref{PV}).
The integral can be evaluated by using the ML prescription as well.
Once we choose $n$ arbitrarily, some loop-integrals 
depending on this spurious pole are ill-defined even under dimensional 
regularization within the PV prescription. For example, there is a scalar 
integral in the gluon self-energy diagram: 
\begin{eqnarray}
\int \frac{d^{D}k}{(2\pi)^{D}} 
\;\frac{1}{\left[(l+k)^2+i\epsilon\right]\; k\cdot n},
\label{ill}
\end{eqnarray}
where $k$ is loop momentum. 
If we use the PV prescription without any further assumption, 
the integral (\ref{ill}) is proportional to the ill-defined gamma 
function $1/\Gamma(0)$ even though we use the regularized dimension 
$D=4-2\epsilon$. Let us call this by the integral having the {\it ill-defined} 
spurious pole. In the ML prescription, the integral (\ref{ill}) 
becomes a well-defined one.  

However, it is dangerous to impose an {\it ad hoc}  prescription 
for these spurious poles. Once we employ such a prescription, we must check 
renormalizability and unitarity, the cancelation of infrared (IR) 
divergence, case by case. Fortunately, in the case of the fragmentation 
function calculation, there is a possibility to fix the spurious pole structure
by using gauge invariance, so we do not have to depend on a specific 
prescription.
In covariant gauges such as the Feynman gauge, there is no such problem.
The spurious pole from the gluon propagator in the light-cone gauge 
is transformed into the propagator of the eikonal line in the Feynman
gauge. And the $i\epsilon$ sign for the eikonal line propagator is 
well defined. If we calculate a gauge-invariant quantity in the two gauges
and compare the two results, we may fix the spurious pole ambiguity in 
the light-cone gauge.

As an example, let us consider the next-to-leading order calculation of 
a fragmentation function.
The natural choice of the light-cone vector $n$ for defining a fragmentation 
function is $n=(1,0_\perp,-1)$, which has vanishing transverse components 
relative to the daughter-particle momentum $p=(p_0,0_\perp,p_N)$. 
In general, any fragmentation function is expressed in terms of the scalar
products among vectors $p$, $n$, and $\bar{n}=(1,0_\perp,1)$.
Since the transverse dependence is integrated out
in the fragmentation function and the daughter particle is on its mass shell, 
the conjugate light-cone vector $\bar{n}$ is no longer 
independent of $p$ and $n$. Therefore, the fragmentation function in the 
light-cone gauge becomes dependent only on $p$ and $n$ even if we use
the ML prescription.  In this case,
the quark wavefunction renormalization constant $Z_Q=1+\delta Z_Q$ and
the gluon propagator correction factor $\Pi$ in one-loop level are given by 
\begin{eqnarray}
\delta Z_Q^{\rm LC}&=&
i\frac{16\pi\alpha_s\mu^{2\epsilon}}{3}
\bigg[\left(2-N\right)I_{AD} + p^2 I_{ADD}
+ 2 p\cdot n I_{BCD}\bigg],
\label{Z:LC}
\\
\delta Z_Q^{\rm F}&=&
i\frac{16\pi\alpha_s\mu^{2\epsilon}}{3}
\bigg[\left(4-N\right)I_{AD} + p^2 I_{ADD} \bigg],
\label{Z:F}
\\
\Pi^{\rm LC}&=&
-i\;6\pi\alpha_s\mu^{2\epsilon}
\bigg\{
\bigg[7+\frac{1}{N} - \frac{2n_f}{3}\left(1-\frac{1}{N}\right)\bigg]\;I_{AB}
- 8p\cdot nI_{ABC}
\bigg\},
\label{Pi:LC}
\\
\Pi^{\rm F}&=&
-i\;6\pi\alpha_s\mu^{2\epsilon}
\bigg[3+\frac{1}{N} - \frac{2n_f}{3}\left(1-\frac{1}{N}\right)\bigg]I_{AB},
\label{Pi:F}
\end{eqnarray}
where $n_f$ is the number of light quark flavors.
The divergences are regularized using dimensional regularization with
spatial dimensions $N=3-2\epsilon$.
Note that $p$ denotes the momentum of the gluon for $\Pi$ and
twice that of the quark momentum for $Z_Q$.
The superscripts F and LC are used for Feynman gauge and light-cone 
gauge, respectively. 
The scalar integrals $I_{AB\cdots}$ are given in Appendix A. 
The constants (\ref{Z:F}) and (\ref{Pi:F}) for the Feynman gauge have no
light-cone dependent integral while the light-cone-gauge counterparts
(\ref{Z:LC}) and (\ref{Pi:LC}) have such integrals as 
$I_{BCD}$ and $I_{ABC}$.
The integral $I_{ACD}$ does not appear in (\ref{Z:LC})-(\ref{Pi:F}). 
In the light-cone gauge, $I_{ACD}$ appears in the vertex
correction factor (\ref{V:LC}).  
In the Feynman gauge, the light-cone dependent integrals appear only in the 
term (\ref{DELTA:F}) which involves the gluon coupling to the eikonal line.
One must be careful about the $i\epsilon$ prescription
for the light-cone dependent denominator $C$ defined in (\ref{eq:C}). 
In the Feynman gauge calculation, the sign of $\epsilon$ is fixed since
the propagators of both the gluon and the eikonal line are well defined. 
But we do not assume that the same sign is also valid for the quantities 
(\ref{Z:LC}) and (\ref{Pi:LC}) in the light-cone gauge. 
The appropriate $i\epsilon$ prescription in the light-cone gauge
can be fixed by comparing results for a gauge-invariant quantity
calculated in the two gauges. 

The light-cone dependent scalar integrals $I_{ABC}$, $I_{ACD}$ and $I_{BCD}$ 
appearing in the light-cone gauge correction factors,
possess important features.
First, they do not have the {\it ill-defined} spurious pole,
even though they have {\it ambiguous} spurious poles.
The integral (\ref{ill}) with the {\it ill-defined} spurious pole 
cancels in the gluon self energy correction factor $\Pi^{\rm LC}$ in 
(\ref{Pi:LC}), so the integrals in (\ref{Z:LC}) and (\ref{Pi:LC})
can be dimensionally regularized within the naive PV prescription (\ref{PV})
as well as the ML prescription. 
The reason why the ill-defined spurious pole does not appear in the light-cone
gauge calculations in (\ref{Z:LC}) and (\ref{Pi:LC}) is that we have 
chosen the light-like vector $n$ with vanishing transverse components relative 
to the momentum of the daughter particle.
Furthermore, their values are independent of the sign of the $i\epsilon$ 
in the definition (\ref{eq:C}). 
Since the integral is invariant under the inversion 
$l\to -l$ and $I_{XYC}|_{p\to -p}=-I_{XYC}$, where $X$ and $Y$ are 
$A$, $B$ or $D$ defined in Appendix A,
it is trivial to obtain the relations :
\begin{eqnarray}
I_{ABC}=I_{ABC^*},\quad
I_{ACD}=I_{AC^*D},\quad
I_{BCD}=I_{BC^*D}.
\label{LCI}
\end{eqnarray}
where the definition of the integral $I_{XYC^*}$ is the same as that of the 
integral $I_{XYC}$ except for the fact that $C=(p-l)\cdot n+i\epsilon$ 
is replaced by its complex conjugate $C^*=(p-l)\cdot n-i\epsilon$.
At least in this case, the values of the scalar integrals agree with 
those evaluated by using the PV prescription : 
\begin{eqnarray}
I_{XYC}&\to&\frac{1}{2}\left(I_{XYC}+I_{XYC^*}\right).
\end{eqnarray}
The independence on the sign in front of the $i\epsilon$ in $C$
might be accidental. 
If we use the ML prescription, each light-cone-dependent integral 
in (\ref{LCI}) has ultraviolet (UV) and IR structures 
which are different from those shown in Appendix A.
Effectively, the ML prescription transforms a double pole into an IR pole
and makes the integral satisfy naive power counting rules.
Note that $Z_Q$ and $\Pi$ are gauge dependent.
The values in the Feynman gauge agree with well-known ones that can be 
found, for example, in Ref. \cite{B-L}.
The result using the ML prescription is known only for the UV poles.
We have full agreement in the UV poles if we use the ML prescription :
the gluon propagator correction term is proportional to the QCD beta function
as $\Pi=(33-2n_f)\alpha_s/(12\pi\epsilon_{\rm UV})$ \cite{ML-L,PI}
and $\delta Z^{\rm LC}_Q=\alpha_s/(3\pi\epsilon_{\rm UV})$ \cite{ZV}.
All of them are listed , for example,  in Ref. \cite{ML-2loop}.
Since we will use gauge invariance to determine the spurious pole structure,
we do not proceed with the prescription dependence
further. A thorough study of the application of the ML prescription to  
this problem will be presented elsewhere \cite{Lee}.

\section{Collins-Soper definition and the light-cone gauge}
\label{sec:CS}

The fragmentation function $D_{g \to H}(z,\mu)$
gives the probability that a gluon produced in a hard-scattering process
involving momentum transfer of order $\mu$ decays into a hadron $H$
carrying a fraction $z$ of the gluon's longitudinal momentum.
This function can be defined in terms of the matrix element of a bilocal 
operator involving two gluon field strengths in a light-cone gauge
\cite{C-F-P}.  In Ref. \cite{C-S},
Collins and Soper introduced a gauge-invariant definition of the gluon 
fragmentation function that involves the matrix element 
of a nonlocal operator consisting of two gluon field strengths and 
eikonal operators.  One advantage of this definition is that it 
avoids subtleties associated with products of singular distributions.  
The gauge-invariant definition is also advantageous for explicit 
perturbative calculations, because it allows the calculation of 
radiative corrections to be simplified by using the Feynman gauge.   

The gauge-invariant definition 
of Collins and Soper for the gluon fragmentation function for
splitting into a hadron $H$ is  
\begin{eqnarray}
D_{g \to H}(z,\mu) &=&
{(-g_{\mu \nu})z^{N-2} \over 16\pi(N-1) k^+}
\int_{-\infty}^{+\infty} dx^- e^{-i k^+ x^-}
\nonumber\\
&&\times
\langle 0 | G^{+\mu}_c(0)
{\cal E}^\dagger(0^-)_{cb} \; {\cal P}_{H(z k^+,0_\perp)} \;
{\cal E}(x^-)_{ba} G^{+ \nu}_a(0^+,x^-,0_\perp) | 0 \rangle\;.
\label{D-def}
\end{eqnarray}
The operator ${\cal E}(x^-)$ in (\ref{D-def}) is an eikonal operator
that involves a path-ordered exponential of gluon field operators along
a light-like path:
\begin{equation}
{\cal E}(x^-)_{ba} \;=\; {\rm P} \exp
\left[ +i g \int_{x^-}^\infty dz^- A^+(0^+,z^-,0_\perp) \right]_{ba},
\label{E}
\end{equation}
where $A^\mu(x)$ is the matrix-valued gluon field in the adjoint
representation:  $[ A^\mu(x) ]_{ac} = if^{abc} A_b^\mu(x)$.
The operator ${\cal P}_{H(p^+,p_\perp)}$ in (\ref{D-def})
is a projection onto states that, in the asymptotic future, contain 
a hadron $H$ with momentum $p = (p^+,p^-=(m_H^2+ p_\perp^2)/p^+,p_\perp)$, 
where $m_H$ is the mass of the hadron.
The hard-scattering scale $\mu$ in (\ref{D-def})
can be identified with the renormalization scale of the nonlocal operator.
The prefactor in the definition (\ref{D-def})
has, therefore, been expressed as a function of the number 
of spatial dimensions $N=3-2\epsilon$.
This definition is particularly useful when we 
use dimensional regularization to regularize ultraviolet divergences.  
If the production process of the hadron $H$ can be described by perturbation
theory, one can use the definition (\ref{D-def}) to calculate the fragmentation
function $D_{g\to H}(z,\mu)$ as a power series in $\alpha_s$.
In Ref.\cite{C-S}, complete sets of Feynman rules for this perturbative
expansion for quark and gluon fragmentation functions are given.
By inserting the eikonal operator (\ref{E}), the operator consisting of two 
gluon fields with different locations becomes gauge invariant.
At higher order in $\alpha_s$, there are numerous diagrams
which have gluons coupled to the eikonal lines.
In the light-cone gauge, the contribution of the eikonal operator disappears
since the gluon decouples from the eikonal line.
Therefore, there is a great reduction in the number of Feynman diagrams.
On the other hand, the spurious pole contribution of the gluon propagator
appears in the light-cone gauge. 
However, the gauge invariance of this definition (\ref{D-def})
provides the  gauge transformation of the eikonal line contribution in the 
Feynman gauge into the spurious pole contribution in the light-cone gauge.
By comparing the final results for the gauge-invariant quantity  
$D_{g\to H}(z,\mu)$ from the two gauges, the spurious pole coming from 
the gluon propagator in the light-cone gauge can be fixed unambiguously.
\section{Application}
\label{sec:CAL}
One remarkable example of a fragmentation phenomenon is 
charmonium production at the Fermilab Tevatron.
The production rate of a heavy quarkonium depends on the cross section
of a heavy quark pair $Q \overline{Q}$  with small relative momentum.
In high-energy $p \bar p$ collisions, the gluon production rate is dominant
and the inclusive production of a heavy quark pair
$Q \overline{Q}$ via subsequent decay of this almost on-shell gluon
is enhanced by the gluon propagator  \cite{B-Y:S1}. 
Furthermore, at leading order
in $\alpha_s$, such a $Q \overline{Q}$ pair created by the virtual gluon
is dominated by a color-octet $^3S_1$ state \cite{B-F}.
The color-octet $^3S_1$ contribution has particular phenomenological importance.
Braaten and Yuan showed that,
in the gluon fragmentation function for splitting
into triplet $P$-wave states, the infrared divergence in the short-distance
coefficient of the color-singlet matrix element 
$\langle {\cal O}_1(^3P_J) \rangle$ can be avoided  by including the 
color-octet $^3S_1$ term \cite{B-Y:P}.
The production rate of direct $J/\psi$ and $\psi'$ at large $p_T$
at the Tevatron \cite{CDF-unpol} is explained by Braaten and Fleming by 
introducing this $\langle {\cal O}_8(^3S_1) \rangle$ term \cite{B-F}.

In the NRQCD factorization formalism \cite{B-B-L},
the fragmentation function $D(z,\mu)$ for a parton splitting
a heavy quarkonium is expressed as a linear combination of NRQCD matrix
elements, which can be regarded as phenomenological parameters.
The corresponding short-distance factors depend on $z$ and are calculable
in perturbation theory.
Most of the phenomenologically relevant short-distance factors
have been calculated to leading order in $\alpha_s$.
They all begin at order $\alpha_s^2$ 
or higher\footnote{for the color-singlet $^3S_1$ channel, the short-distance 
factor begins at order $\alpha_s^3$}, with the exception 
of the color-octet $^3S_1$ term in the gluon fragmentation function,
which begins at order $\alpha_s$. 
Since  the color-octet $^3S_1$ term dominates the high-$p_T$ 
gluon fragmentation phenomena in heavy quarkonium production,
the next-to-leading order correction of order $\alpha_s^2$ 
to this term is particularly important.

As an application, we consider the next-to-leading order
correction to the color-octet $^3S_1$ gluon fragmentation function
for heavy quarkonium $H$. Since there is a discrepancy between CDF data
and the leading-order prediction of the prompt $J/\psi$ polarization at 
large $p_T$ where the gluon fragmentation contribution is important
\cite{CDF-pol,B-K-L-1,B-K-L-2,L}, the full NLO
calculation of the polarized heavy quarkonium production rate is needed too.
Unfortunately, there are two different results for the color-octet
$^3S_1$ term \cite{Ma-2,B-L}. 
Therefore, it is worth while to calculate this important function
in an independent way.
Since both previous calculations employed the Feynman gauge,
we shall present our results in the light-cone gauge.
In order to determine the appropriate prescription for the spurious poles, 
we use the result from the Feynman gauge. By comparing the two intermediate 
results before the evaluation of the light-cone dependent integrals,
we fix the sign of $i\epsilon$ in the spurious pole in the light-cone gauge. 

We use the same conventions as those presented in Ref. \cite{B-L}. 
We do not reproduce the description on the theoretical background of 
the fragmentation function for heavy quarkonium production in NRQCD 
factorization formalism which is well explained in Ref. \cite{B-L}.
Based on the NRQCD factorization formalism \cite{B-B-L},
the fragmentation function is written in a factorized form \cite{B-L}:
\begin{eqnarray}
D_{g \to H}(z) &=&
\left[ (N-1) d_T(z) + d_L(z)  \right]
\langle {\cal O}_8^H(^3S_1) \rangle,
\label{D-sum}
\end{eqnarray}
where $d_T$ and $d_L$ are the short-distance coefficients for
the transverse and longitudinal contributions and
$\langle {\cal O}_8^H(^3S_1) \rangle$ is the color-octet $^3S_1$
matrix element  defined in Ref. \cite{B-B-L}.

\begin{figure}
\begin{center}
\epsfig{file=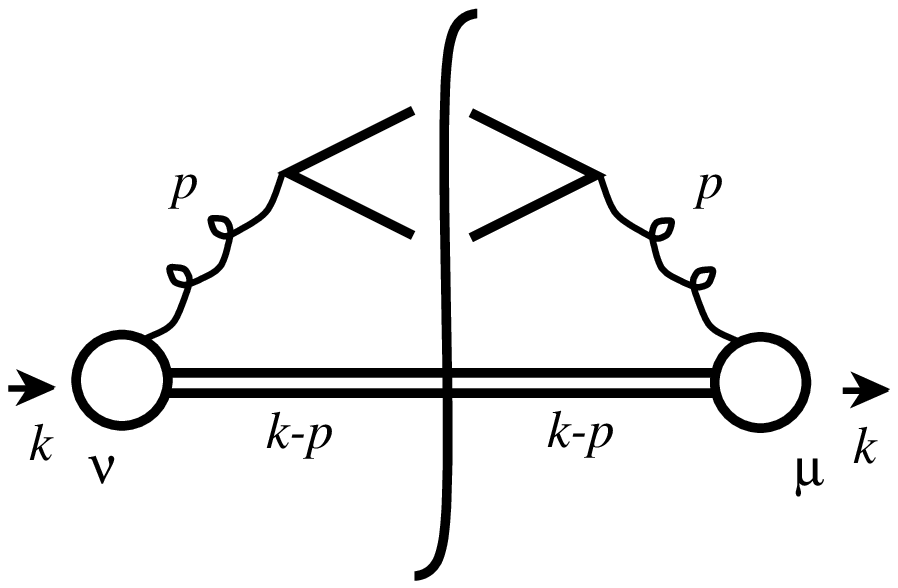, height=30 ex}
\end{center}
\caption{Leading order Feynman diagram for $g \to Q\overline{Q}$.}
\label{2body0}
\end{figure}

There is only one lowest-order diagram in both 
Feynman and light-cone gauge, which is shown in Fig.~\ref{2body0}.
The circles connected by the double pair of lines represent the
nonlocal operator consisting of the gluon field strengths and
the eikonal operators in the definition (\ref{D-def}).
The momentum $k = (k^+,k^-,k_\perp)$ flows into the circle on the left
and out of the circle on the right.
The cutting line represents the projection onto states which, in the
asymptotic future, include a $Q \overline{Q}$ pair with total momentum
$p = (z k^+, p^2/(z k^+), 0_\perp)$.
The appearance of the diagrams for both gauges is the same 
in this order, since the circle should emit a gluon.
With the Feynman rules of Ref. \cite{C-S} and following the method of 
extracting the short-distance coefficients of the fragmentation function 
in Ref. \cite{B-L},
we can read off the order-$\alpha_s$ terms in the short-distance functions
$d_T(z)$ and $d_L(z)$ as 
\begin{eqnarray}
d_T^{(\rm LO)}(z) &=&
\frac{\pi\alpha_s\mu^{2\epsilon}}{8N(N-1)m_Q^3}
\;\delta(1-z),
\label{d1-T}
\\
d_L^{(\rm LO)}(z) &=& 0.
\label{d1-L}\end{eqnarray}
We have neglected the relative momentum of the heavy quark 
in the $Q \overline{Q}$ rest frame so that the invariant mass of the
pair is $p^2 = 4 m_Q^2$.
The LO results (\ref{d1-T}) and (\ref{d1-L}) agree with previous calculations 
in the Feynman gauge \cite{B-L,B-C:TEdr}.

\begin{figure}
\begin{center}
\epsfig{file=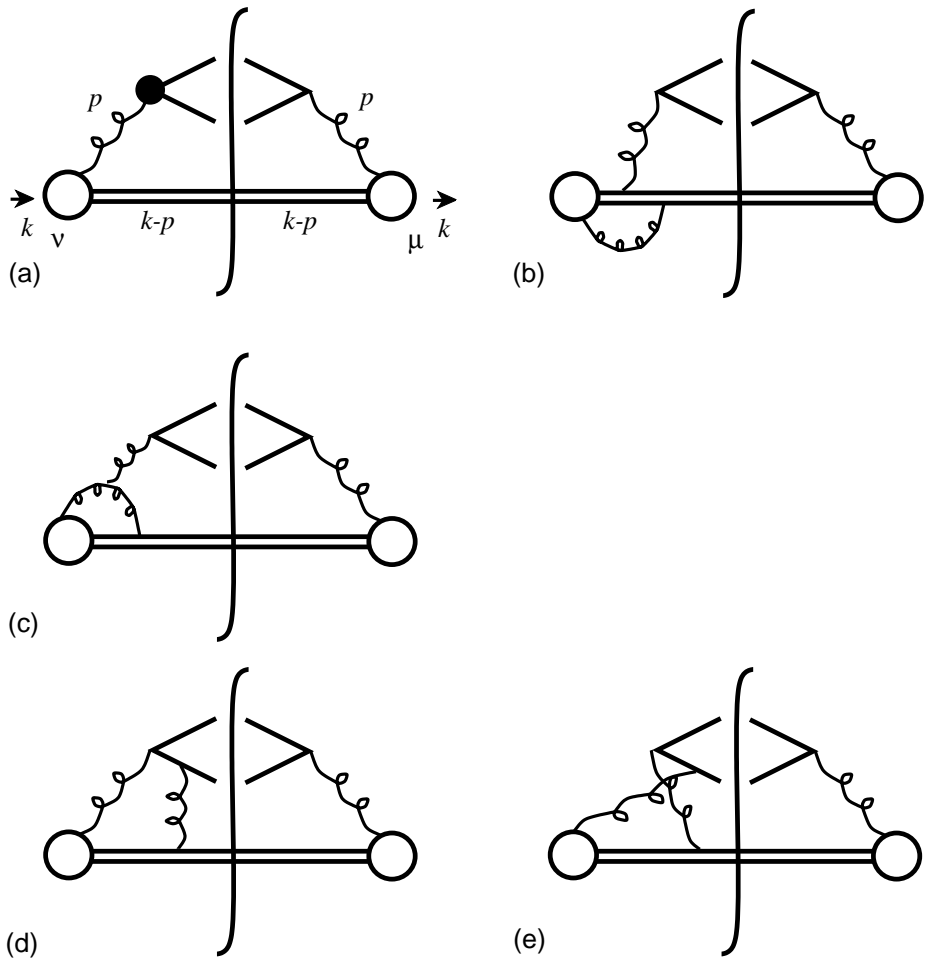, height=70ex}
\end{center}
\caption{The Feynman diagrams of order $\alpha_s^2$ for $g \to Q\overline{Q}$
with $Q\overline{Q}$ final states.  There are additional contributions
from the complex-conjugate diagrams.}
\label{2bodya}
\end{figure}

\begin{figure}
\begin{center}
\epsfig{file=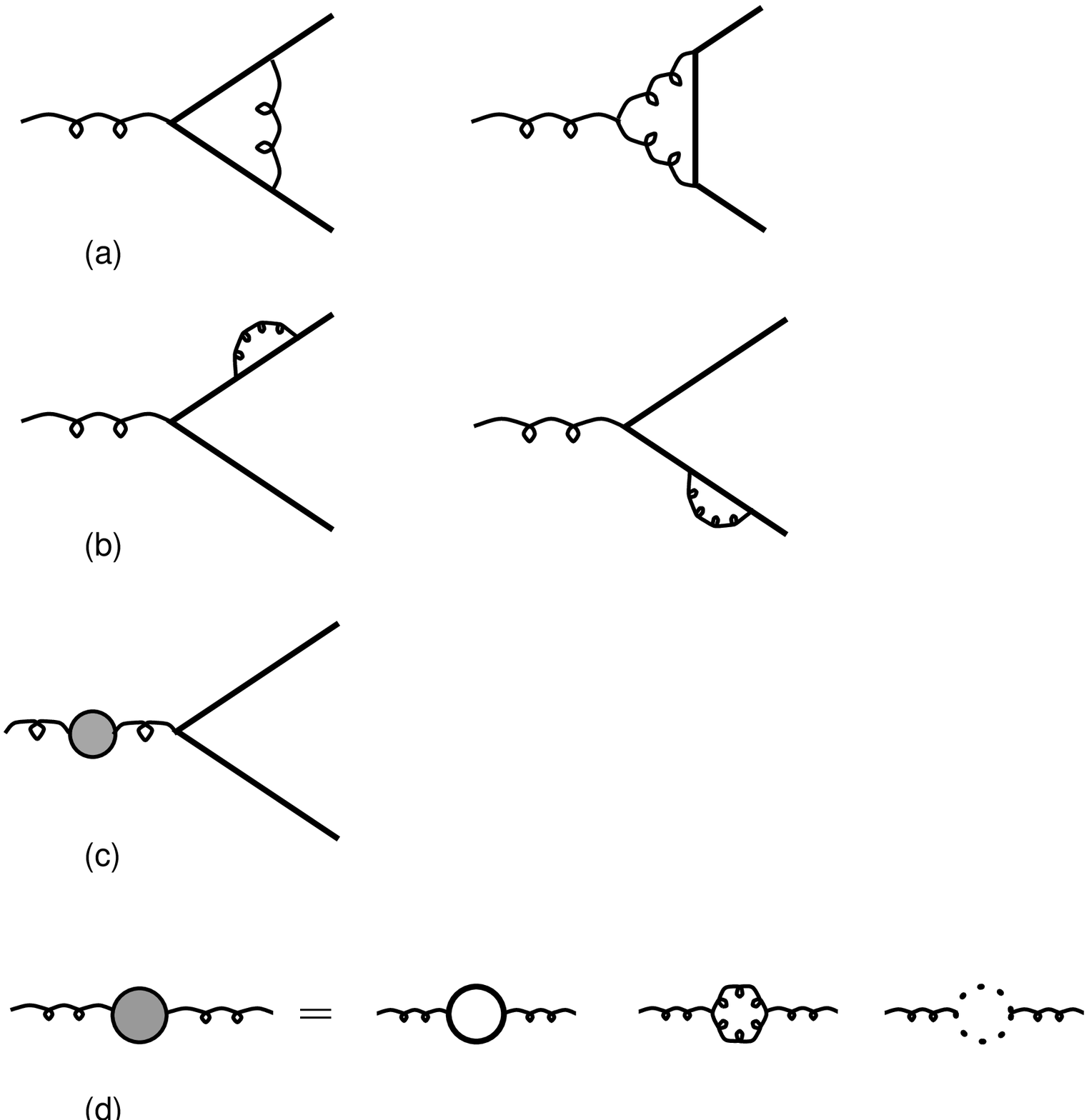, height=70ex}
\end{center}
\caption{One loop correction diagrams for
$g^*\to Q\overline{Q}$.
        }
\label{2onlya}
\end{figure}

The Feynman diagrams for the 
fragmentation function for $g \to Q\overline{Q}$ at order $\alpha_s^2$
consist of virtual corrections, for which the final state is $Q\overline{Q}$,
and real-gluon corrections, for which the final state is $Q\overline{Q}g$.
The diagrams with virtual-gluon corrections to the left of the cutting line
are shown in Fig.~\ref{2bodya}.  The black blob in  Fig.~\ref{2bodya}(a)
includes the vertex corrections and propagator corrections shown in 
Fig.~\ref{2onlya}.
In the Feynman gauge, only the diagram in Fig.~\ref{2bodya}(b) vanishes,
because the gluon attached to the eikonal line gives a factor of $n^\mu$.
On the other hand, all the diagrams except for Fig.~\ref{2bodya}(a) 
vanish in the light-cone gauge.
If we use the threshold-expansion method of Braaten and Chen 
\cite{B-C:TE}, we can simplify the structure of the expression 
without employing the projection method. With the threshold expansion, 
we can keep the full structure of color and spin.  
Here we utilize the dimensionally regularized threshold expansion method 
of  Braaten and Chen \cite{B-C:TEdr,B-C:Pdecay}.
With the Dirac equation and the usual methods for reducing tensor integrals 
into scalar integrals, we factorize each virtual correction 
diagram into the leading order diagram in Fig.~\ref{2body0}, times 
a multiplicative factor.
In the light-cone gauge, the ghost decouples since its coupling to the
gluon is orthogonal to the gluon propagator (\ref{Gpro:LC}), 
so the gluon propagator correction factor shown in Fig. 3(d) does not 
have ghost contribution.

The virtual corrections contribute only to the transverse short-distance
function $d_T(z)$ defined in \cite{B-L}:
\begin{eqnarray}
d_T^{\rm (virtual)}(z) = d_T^{(\rm LO)}(z)\times
\;2\;{\rm Re}\bigg[\Lambda + \Pi + \delta Z_Q    + \Delta \bigg],
\end{eqnarray}
where $\delta Z_Q$ and $\Pi$ are defined in (\ref{Z:LC})--(\ref{Pi:F})
and $\Lambda$ is the vertex correction factor.
The contribution from the remaining diagrams shown
in Fig. 2 (b)-(e), which have gluon couplings to the eikonal lines,
is expressed as $\Delta$.
Their values are expressed in terms of one-loop scalar integrals:
\begin{eqnarray}
\Lambda^{\rm LC}&=&
i\;
\frac{2\pi\alpha_s\mu^{2\epsilon}}{3}
\bigg[9\left(7+\frac{ 1}{N}     \right)\;I_{AB}
     + \left(N+\frac{18}{N} - 67\right)\;I_{AD}
     -p^2 I_{AAD}
\nonumber\\
&&\quad\quad\quad\quad\quad
     +2\;p\cdot n\; \left(9I_{ACD} + I_{BCD}-36I_{ABC}\right)
\bigg],
\label{V:LC}
\\
\Lambda^{\rm F}&=&
i\;
\frac{2\pi\alpha_s\mu^{2\epsilon}}{3}
\bigg[ 9\left(1+\frac{ 1}{N}     \right)I_{AB}
     +  \left(N+\frac{18}{N} - 47\right)I_{AD}
     -p^2 I_{AAD}
\bigg],
\label{V:F}
\\
\Delta^{\rm LC}&=&0,
\\
\Delta^{\rm F}&=&
i\;
12\;{\pi\alpha_s\mu^{2\epsilon}}
\bigg[I_{AB}-2I_{AD}
      + p\cdot n\left(I_{ACD}+I_{BCD}\right) \bigg].
\label{DELTA:F}
\end{eqnarray}
The explicit value of the vertex correction factor $\Lambda^{\rm F}$ in the
Feynman gauge shown in (\ref{V:F}) agrees with the result in Ref. \cite{B-L}.
The UV dependence of the vertex correction factor $\Lambda^{\rm LC}$ 
in the light-cone gauge shown in (\ref{V:LC}) agrees with the result using 
the ML prescription in Refs. \cite{ZV,ML-2loop} where only the UV contribution
is given :
$\Lambda^{\rm LC}_Q=-\delta Z^{\rm LC}_Q=-\alpha_s/(3\pi\epsilon_{\rm UV})$.
The integral $I_{AAD}$ has a Coulomb singularity  
as well as a logarithmic IR divergence due to the exchange of a gluon between 
the on-shell heavy quark and anti-quark. 
Dimensional regularization puts power infrared divergence like the
Coulomb singularity to zero, so only the logarithmic IR divergence remains in
the integral $I_{AAD}$. Then the integral is effectively expressed by $I_{ADD}$
via the equation $I_{ADD}=(N-4)I_{AAD}$.
It is important to notice that various correction factors in 
the Feynman and the light-cone gauge involve different combinations of the
same scalar integrals. 
Straight-forward sums for both gauges produce a common result
\begin{eqnarray}
d_T^{\rm (virtual)}(z) = d_T^{(\rm LO)}(z)
\frac{4\pi\alpha_s}{3}
\;{\rm Re}\bigg\{i\left[
            -\left(7N-\frac{18}{N}+51\right) I_{AD}
      +  6n_f\left(1 -\frac{ 1}{N}   \right) I_{AB}
\right.
\nonumber\\
\left.
     + 18\;p\cdot n\left(I_{ACD}+I_{BCD}\right)
     + p^2\left(8I_{ADD}-I_{AAD}\right)
\right]
\bigg\}.
\label{dT-sum}
\end{eqnarray}
Thus the non-vanishing contributions from the gluon
coupling to the eikonal line in the Feynman gauge, $\Delta^{\rm F}$, is 
simply distributed to other correction factors in the light-cone gauge 
via additional gluon propagator terms.

Since gauge invariance holds for both the virtual and the real-gluon corrections
separately, the equality of the virtual corrections in the Feynman and the 
light-cone gauge is a consequence of gauge invariance.
As we commented in the previous section, the light-cone dependent integrals
in the Feynman gauge result have no ambiguities form spurious poles.
On the other hand, we have not fixed the sign of the $i\epsilon$ in the 
spurious pole of the integrals which are obtained in the light-cone gauge.
Since we have found exact agreement between the two results in the two gauges,
we may simply use the values obtained from the Feynman gauge calculation.
Note that the integral $I_{ABC}$ disappears in (\ref{dT-sum}), so the only 
light-cone dependent integrals that survive are  $I_{ACD}$ and $I_{BCD}$.
The values of these integrals are independent of the sign of the $i\epsilon$
in the definition of $C$ in (\ref{eq:C}).  The expansion of (\ref{dT-sum}) 
in $\epsilon$ reproduces the result of Braaten and Lee
\cite{B-L}:
\begin{eqnarray}
&&d_T^{\rm (virtual)}(z) = d_T^{(\rm LO)}(z)\;
{\alpha_s \over \pi}  \left( {\pi \mu^2 \over m_Q^2} \right)^\epsilon
\nonumber\\
&&
\quad\quad
 \times
\left[  {3(1-\epsilon) \over 2}{\Gamma(1+\epsilon) \over 
         \epsilon_{\rm UV} \epsilon_{\rm IR}} 
	+ \beta_0 {\Gamma(1+\epsilon) \over \epsilon_{\rm UV}}
	+ {177-10n_f \over 18} - {\pi^2 \over 2} 
	+ 8 \ln 2 + 6 \ln^2 2  \right],
\label{d-virtual}
\end{eqnarray}
where $\beta_0=(33-2n_f)/6$.

\begin{figure}
\begin{center}
\epsfig{file=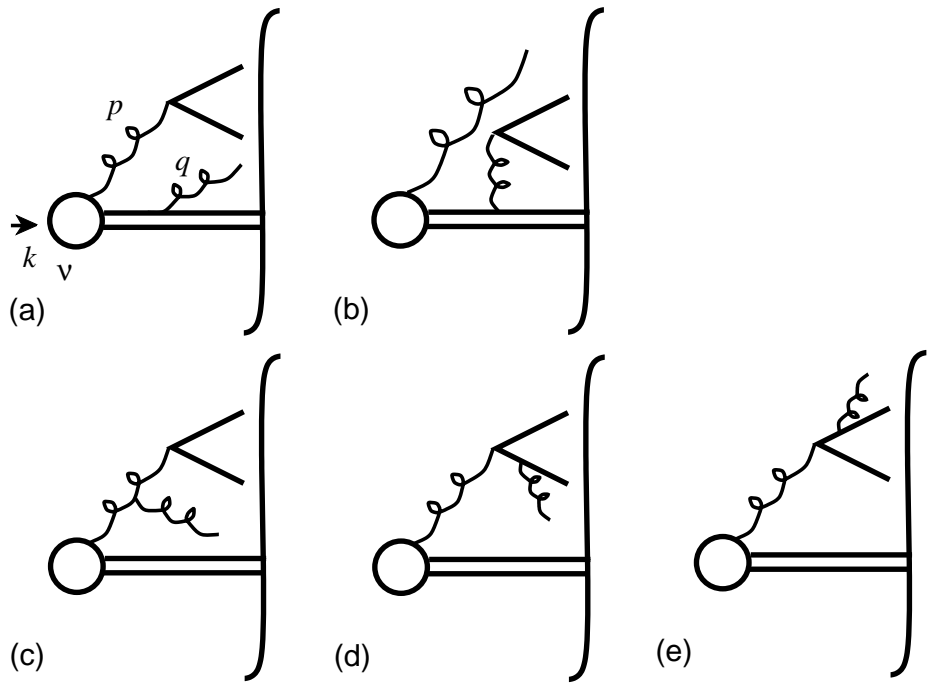, height=50ex}
\end{center}
\caption{
The Feynman diagrams of order $\alpha_s^2$ for $g \to Q\overline{Q}$
with $Q\overline{Q}g$ final states.  There are a total of 25 diagrams,
but only the left halves of the diagrams are shown.}
\label{3body}
\end{figure}

The Feynman diagrams for the real-gluon corrections to the fragmentation
function for $g\to Q\overline{Q}$ can also be calculated in both 
gauges. We draw the 5 left-half diagrams only, which must be multiplied by
their complex conjugates to give a total of 25 diagrams.
The real-gluon correction is a tree-level calculation.
Therefore, there is  no spurious pole problem.
In the Feynman gauge, all 25 diagrams contribute, while only
9 diagrams in the light-cone gauge. In the latter gauge,
diagrams \ref{3body}(a) and \ref{3body}(b) vanish.
The real-gluon correction contribution 
is also gauge invariant.
Employing either gauge,
we reproduce the real correction contribution given in Ref \cite{B-L}
before the phase-space integral is performed:
\begin{eqnarray}
d_T^{\rm  (real)}(z)&=&
\frac{\pi\alpha_s\mu^{2\epsilon}}
{8N(N-1)m_Q^3}
\times
\frac{3\alpha_s}{\pi\Gamma(1-\epsilon)}
\left(
\frac{\pi \mu^2}{m_Q^2}
\right)^\epsilon
\left(1-\frac{1}{z(1-z)}\right)^2
\int_{(1-z)/z}^\infty
dx
\frac{t^{1-\epsilon}}{x^2}\;,
\label{eq:epsIR}
\\
d_L^{\rm  (real)}(z)&=&
\frac{\pi\alpha_s\mu^{2\epsilon}}
{8Nm_Q^3}
\times
\frac{3\alpha_s}{\pi\Gamma(1-\epsilon)}
\left(
\frac{\pi \mu^2}{m_Q^2}
\right)^\epsilon
\left(\frac{1-z}{z}\right)^2
\int_{(1-z)/z}^\infty
dx
\frac{t^{-\epsilon}}{x^2}\;,
\end{eqnarray}
where $t=(1-z)(zx+z-1)$, $x=2q\cdot p/p^2$, 
$q$ is the final-state gluon momentum, and $p$ is the 
$Q\overline{Q}$ momentum. 
The final results for the real-gluon correction contribution of Braaten and 
Lee are straight-forwardly reproduced:
\begin{eqnarray}
d_T^{\rm (real)}(z)
&=&
\frac{\pi\alpha_s\mu^{2\epsilon}}
{8N(N-1)m_Q^3}
\times
\frac{\alpha_s}{\pi}
\;
\left(\frac{\pi\mu^2}{m_Q^2}\right)^\epsilon \Gamma(1+\epsilon)
\nonumber\\
&&\times
\;\bigg[
-\frac{3(1-\epsilon)}{2\epsilon_{\rm UV}\epsilon_{\rm IR}}
\delta(1-z)
+\frac{3(1-\epsilon)}{\epsilon_{\rm UV}}
\left(
\frac{z}{(1-z)_+}+\frac{1-z}{z}+z(1-z)
\right)
\nonumber\\
&&
\quad\;\;\;
  -\frac{6}{z}\left(\frac{\ln(1-z)}{1-z}\right)_+
+6(2-z+z^2)\ln(1-z)
\bigg]\;,
\label{d-real}
\\
d_L^{\rm (real)}(z)
&=&
\frac{\pi\alpha_s}
{8Nm_Q^3}
\times
\frac{3\alpha_s}{\pi}
\frac{1-z}{z}\;.
\label{dL-real}
\end{eqnarray}

The infrared divergence cancels after summing the real and virtual 
correction contributions shown in (\ref{d-virtual}) and (\ref{d-real}).
Employing the $\overline{\rm MS}$ scheme, $\alpha_s$ and the operator are
renormalized as in Ref. \cite{B-L}. After renormalization, 
the final answers for $d_T(z)$ and $d_L(z)$ of Braaten and Lee 
\cite{B-L} are reproduced:
\begin{eqnarray}
d_T(z,\mu)
=
\frac{\pi\alpha_s(\mu)}{48m_Q^3}
\;
&\bigg\{&
\;
\delta(1-z)+\frac{\alpha_s(\mu)}{\pi}
\bigg[
A(\mu)\delta(1-z)
+\left(
 \ln\frac{\mu}{2m_Q}-\frac{1}{2}
 \right)P_{gg}(z)
\nonumber\\
&&+6(2-z+z^2)\ln(1-z)
  -\frac{6}{z}\left(\frac{\ln(1-z)}{1-z}\right)_+
\;
\bigg]
\bigg\}
\;,
\label{dT-final}
\end{eqnarray}
where the coefficient $A(\mu)$ is
\begin{eqnarray}
A(\mu)=
\beta_0\left(\ln\frac{\mu}{2m_Q}+\frac{13}{6}\right)
+\frac{2}{3}-\frac{\pi^2}{2}+8\ln 2+6\ln^2 2
\;,
\end{eqnarray}
and $P_{gg}(y)$ is the gluon splitting function:
\begin{eqnarray}
P_{gg}(z)=
6\left[
\frac{z}{(1-z)_+}+\frac{1-z}{z}+z(1-z)+\frac{\beta_0}{6}\;\delta(1-z)
\right]\;.
\end{eqnarray}
The transverse term $d_T(z)$ in (\ref{dT-final}) still disagrees 
with that of Ma \cite{Ma-2}.
Our final answer for the longitudinal fragmentation function is 
obtained by setting $\epsilon \to 0$ in (\ref{dL-real}):
\begin{eqnarray}
d_L(z,\mu)
=\frac{\alpha_s^2(\mu)}{8m_Q^3}\;\frac{1-z}{z} \,.
\label{dL-final}
\end{eqnarray}
The longitudinal term, $d_L(z)$ agrees with that of Braaten and Lee \cite{B-L}
as well as that of Beneke and Rothstein\cite{B-R}.
The dependence on the spectroscopic state of the produced quarkonium 
of this fragmentation function can be found in Ref. \cite{B-L}.
\section{Discussion}
\label{sec:DIS}
We have shown how the light-cone gauge can be used to evaluate the 
perturbatively calculable parts of a fragmentation function.
As an application, we tested our method by evaluating
the next-to-leading order correction to 
the color-octet $^3S_1$ term in the gluon fragmentation function. 
We reproduced the recent result of Braaten and Lee \cite{B-L}
which disagrees with that of Ma \cite{Ma-2}.
The light-cone gauge considerably simplifies the calculation procedure for 
both the real and the virtual corrections.
At least at the one-loop level, the spurious pole problem can be resolved. 
This problem does not appear in the real corrections,
because they come from tree-level diagrams, but it does appear in the virtual 
corrections.  The gauge-invariant definition of the fragmentation function 
of Collins and Soper allows us to fix the ambiguities from spurious poles 
in the light-cone gauge by comparing with the result obtained in
the Feynman gauge.
We reduced the virtual correction in the color-octet $^3S_1$ 
fragmentation function in the light-cone gauge to a linear combination
of scalar integrals. After naive cancelations among the scalar integrals,
ignoring the ambiguity from spurious poles,
the correction reduces to scalar integrals that are independent of the sign
of $\epsilon$ in the denominator $k\cdot n+i\epsilon$. Thus the PV prescription
gives the correct answer.  To see if the ML prescription also gives correct 
answer requires explicit calculations of the scalar integrals including
IR and finite terms.
As a byproduct, the renormalization constants in the
light-cone gauge were obtained at one-loop level.  Their UV dependencies agree 
with the previous calculations within the ML prescription. 
They might be useful for 
other calculations, such as the next-to-leading order corrections to other 
fragmentation functions \cite{others}.

Were it not for the problem of the spurious pole, one could reduce a        
large amount of the intermediate calculation by using the light-cone gauge.
If we choose the gauge-fixing vector $n$ with vanishing transverse
components with respect to the momentum of the produced hadron,
the one-loop integrals are remarkably simplified compared to the case
using the ML prescription.
We were able to determine the spurious pole structure 
without depending on a specific prescription by comparing with the result
in the Feynman gauge. Of course, if we have to repeat the entire calculation
using the Feynman gauge to fix the spurious poles, the light-cone gauge will
not save any labor. However,  there are only a small number of integrals
which have the spurious pole problem, and this provides a way to save labor.
First calculate the full contribution in the light-cone gauge in terms of
scalar integrals, without specifying any prescription for the spurious poles.
Then, calculate in the Feynman gauge only those diagrams where 
the eikonal line couples with one or more gluons.
By comparing the two results, we can determine the appropriate prescription
for the spurious poles in the integrals. One restriction of this method is 
that it can only be applied to gauge-invariant quantities.

 The study of high-$p_T$ fragmentation phenomena 
has a significant potential to refine our understanding of hadron physics.
It avoids the nontrivial resummations 
that complicate theoretical predictions for low-$p_T$ hadron processes.
The signature of fragmentation dominance in high $p_T$
charmonium production has been observed in Run I of the Tevatron. 
In Run II of the Tevatron, as well as at LHC and at future colliders, 
there will be much better statistics of high-$p_T$ heavy meson events.
Quantitative predictions for quarkonium production at high $p_T$ 
will require next-to-leading order calculations of all the 
phenomenologically relevant fragmentation functions.
The light-cone gauge may be a powerful tool for carrying 
out these calculations. 

\bigskip
\centerline{\bf Acknowledgements}
\smallskip

The author thanks  Eric Braaten for stressing the importance of this
subject and for his encouragement, and  George Leibbrandt for providing 
helpful information on the ML prescription and for his valuable comments. 
He expresses special gratitude to  Gustav Kramer for his careful reading of 
the manuscript and for useful suggestions. 
Productive discussions with  Bernd A. Kniehl and Choonkyu Lee are also 
acknowledged.
This work was supported in part by the Alexander von Humboldt Foundation
through Research Fellowship No. IV-KOR/1056268,
and by the Deutsche Forschungsgemeinschaft and the Korea Science and
Engineering Foundation through the German-Korean scientific exchange program 
DFG-446-KOR-113/137/0-1.
\renewcommand {\theequation}{\Alph{section}.\arabic{equation}}
\begin{appendix}
\setcounter{equation}{0}
\section{\boldmath
Integral Table
\unboldmath}
\label{sec:appendix}
In this appendix, we present the explicit values of the integrals
encountered in evaluating the virtual-gluon corrections.
Most of them are presented in Ref. \cite{B-L}. But we reproduce here
for completeness.
These integrals have the form
\begin{equation}
I_{AB\cdots} \;=\; \int\frac{d^{N+1}l}{(2\pi)^{N+1}} \frac{1}{AB\cdots}\;,
\end{equation}
where the denominator $AB\cdots$ can be a product of 1, 2, 3, or 4 of
the following factors:
\begin{eqnarray}
A&=&l^2+i\epsilon,\\
B&=&(l-p)^2+i\epsilon=l^2-2l\cdot p+4 m_Q^2+i\epsilon,\\
C&=&(p-l)\cdot n+i\epsilon,
\label{eq:C}
\\
D&=&(l-p/2)^2-m_Q^2+i\epsilon=l^2-l\cdot p+i\epsilon.
\end{eqnarray}
The momentum $p$ is that of a $Q \overline{Q}$ pair with 
zero relative momentum ($p^2 = 4 m_Q^2$) and $n$ is light-like ($n^2 = 0$).  
The integrals $I_A$ and $I_B$ vanish in dimensional regularization.
By symmetry under $p \to l - p$, we have $I_{AD} = I_{BD}$.
Some of the integrals can be reduced to ones with fewer denominators 
by using the identity $A+B-2D=4 m_Q^2$:
\begin{eqnarray}
4 m_Q^2 I_{ABD} &=& 2(I_{AD}-I_{AB}),
\\
4 m_Q^2 I_{ABCD} &=& I_{ACD}+I_{BCD}-2I_{ABC}.
\end{eqnarray}
The independent integrals that need to be evaluated are therefore
\begin{eqnarray}
I_{AB} &=&
\frac{i}{(4\pi)^{2}}
\left( {\pi e^{i\pi} \over m_Q^2} \right)^\epsilon \;
{\Gamma(1+\epsilon) \Gamma^2(1-\epsilon) 
	\over \epsilon_{\rm UV} \Gamma(2-2\epsilon)}\;,
\\
I_{AD}&=&
\frac{i}{(4\pi)^{2}}
\left( {4\pi \over m_Q^2} \right)^\epsilon \;
{\Gamma(1+\epsilon) \over \epsilon_{\rm UV}(1 - 2\epsilon) } \;,
\\
I_{AAD}&=&
\frac{-i}{(4\pi)^{2}(2m_Q^2)}
\left( {4\pi \over m_Q^2} \right)^\epsilon \;
\frac{\Gamma(1+\epsilon)}{\epsilon_{\rm IR}(1+2\epsilon)} \;,
\\
I_{ADD}&=&
\frac{i}{(4\pi)^{2}(2m_Q^2)}
\left( {4\pi \over m_Q^2} \right)^\epsilon \;
{\Gamma(1+\epsilon) \over \epsilon_{\rm IR}} \;,
\\
I_{ABC}&=&
\frac{-i}{(4\pi)^{2}p\cdot n}
\left( {\pi e^{i\pi} \over m_Q^2} \right)^\epsilon \;
{\Gamma(1+\epsilon) \Gamma^2(1-\epsilon) 
	\over \epsilon_{\rm UV} \epsilon_{\rm IR} \Gamma(1-2\epsilon)}
\;,
\\
I_{ACD}&=&
\frac{+i}{(4\pi)^{2}p\cdot n}
\left( {4\pi \over m_Q^2} \right)^\epsilon \;
\frac{\Gamma(1+\epsilon)}{\epsilon_{\rm UV}}\;
\left[ 2 \ln 2 + \epsilon \left( {\pi^2 \over 3} - 6 \ln^2 2 \right)
	+ O(\epsilon^2) \right] \;,
\\
I_{BCD}&=&
\frac{-i}{(4\pi)^{2}p\cdot n}
\left( {4\pi \over m_Q^2} \right)^\epsilon \;
    {\Gamma(1+\epsilon) \over \epsilon_{\rm UV} \epsilon_{\rm IR}} \;.
\end{eqnarray}
The subscripts on the poles in $\epsilon$ indicate whether the divergences
are of ultraviolet or infrared origin.
\end{appendix}


\newpage

\end{document}